\documentclass[MSNbibl,nameyear,dvips]{arxstspdf}
\usepackage{flushend}
\usepackage{stfloats}
\usepackage{graphicx}
\volume{26}
\issue{2}
\pubyear{2011}
\firstpage{291}
\lastpage{295}
\doi{10.1214/11-STS349B}
\referstodoi{10.1214/10-STS349}

\makeatletter
\renewcommand{\b}{\bolds{\beta}}

\def\bsuffix #1{#1}
\makeatother

\begin{document}
\begin{frontmatter}

\title{Discussion of ``Estimating Random Effects via Adjustment for Density Maximization'' by C. Morris and R. Tang}
\runtitle{Discussion}
\pdftitle{Discussion of Estimating Random Effects via Adjustment for Density Maximization by C. Morris and R. Tang}

\begin{aug}
\author[a]{\fnms{P.} \snm{Lahiri}\corref{}\ead[label=e1]{plahiri@survey.umd.edu}}
and
\author[b]{\fnms{Santanu} \snm{Pramanik}\ead[label=e2]{pramanik-santanu@norc.org}}
\runauthor{P. Lahiri and S. Pramanik}

\affiliation{University of Maryland and University of Chicago}

\address[a]{P. Lahiri is Professor, Joint Program in Survey Methodology, University of Maryland, College Park, Maryland 20742, USA \printead{e1}.}

\address[b]{Santanu Pramanik is Survey Statistician, NORC at the University of Chicago, 4350 East West Highway, 8th Floor, Bethesda, Maryland 20814, USA \printead{e2}.}

\end{aug}



\end{frontmatter}

We thoroughly enjoyed reading this excellent
authoritative paper full of interesting ideas, which\break should be
useful in both Bayesian and non-Bayesian inferences.   We first discuss the accuracy of the ADM
approximation to a Bayesian solution in a real-life application
and then discuss how some of the ideas presented in the paper
could be useful in a non-Bayesian setting.

\section*{How Does the ADM Work in a Real Application?}

Although the main objective of this paper is to make inferences on
the high-dimensional parameters or the random effects $\theta_i$,
the authors note that the success of the Bayesian method lies on
the accurate estimation of the shrinkage parameters $B_i$ since
they appear linearly in the expressions for the posterior mean and
posterior variance of $\theta_i$ when the hyperparameters are
known. Thus, we assess the accuracy of the ADM approximation,
given in Section 2.8, relative to the standard first-order Laplace
approximation, in approximating the posterior distribution of the
shrinkage factors for the hierarchical model~(1)--(2). This model,
commonly referred to as the Fay--Herriot model in the small area
literature, was used by Fay and Herriot (\citeyear{fh79}) in order to combine survey
data and different administrative records in producing empirical
Bayes estimates of per-capita income of small places. Since then
the Fay--Herriot model and its variants have been used in various
federal programs such as the Small Area Income and Poverty
Estimates (SAIPE) and the Small Area Health Insurance estimates
(SAHIE) programs of the U.S. Census Bureau.

For purposes of evaluation, we consider the problem of estimating
the proportion of 5- to 17-year-old (related) children in poverty for
the fifty states and the District of Columbia using the same
data set considered by Bell (\citeyear{bell99}).   We choose two years (1993
and 1997) of state-level data from the SAIPE program. In 1993, the
REML estimate of $A$ is positive while in year 1997 it is zero.
The choice of these two years will thus give us an opportunity to
assess the accuracy of the ADM approximations in two different
scenarios.

We assume the standard SAIPE state-level model in which
survey-weighted estimates of the proportions follow the two-level
model given by (1)--(2). The survey-weighted proportions are
obtained using the Current Population Survey (CPS) data with their
variances $V_i$ estimated by a Generalized Variance Function (GVF)
method, following Otto and Bell (\citeyear{ottobell95}), but assumed to be known
throughout the estimation procedure. We use the same state-level
auxiliary variables $x_i$ (a vector of length 5, i.e., $r=5$),
obtained from Internal Revenue Service (IRS) data, food stamp data
and Census data that the SAIPE program used for the problem. We
assume the uniform prior on $\b$ and superharmonic (uniform) prior
on $A$, as used in the Morris--Tang paper.

\begin{figure*}

\includegraphics{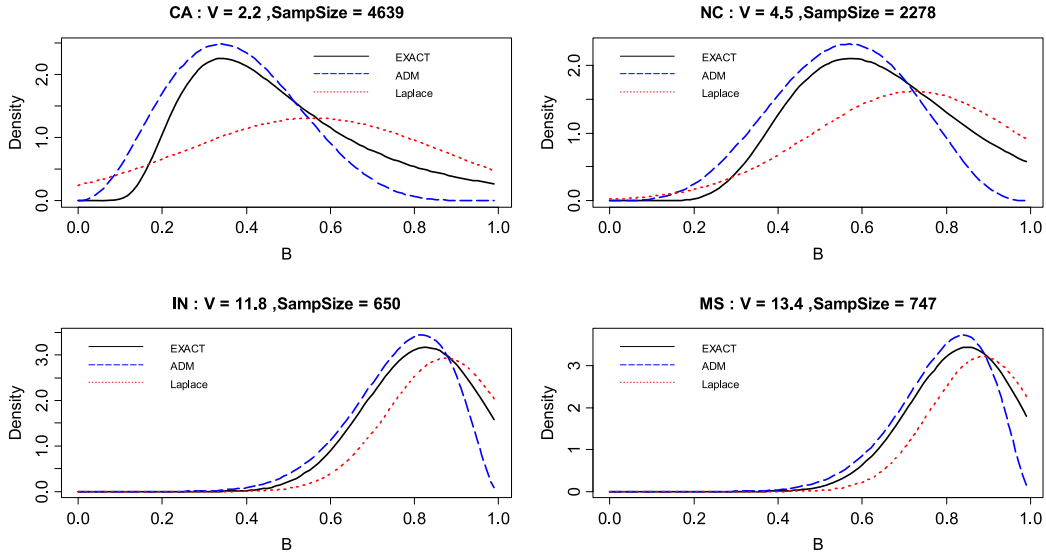}%

\caption{Plot of exact posterior density of $B_i$ along with
approximate densities using SAIPE 1993 state-level data. The four
states in the plot are taken from Bell (\citeyear{bell99}).}
\label{figsaipe93A}
\end{figure*}

\begin{figure*}[b]

\includegraphics{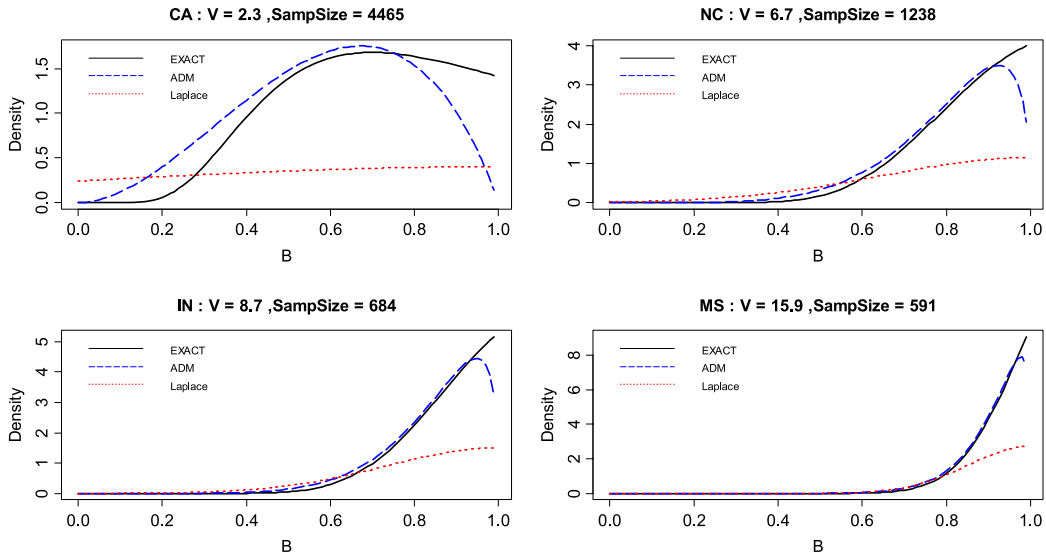}%

\caption{Plot of exact posterior densities of the shrinkage
factors $B_i$ along with approximate densities using SAIPE 1997
data; the four states in the plot are taken from Bell (\citeyear{bell99}).}
\label{figsaipe97A}
\end{figure*}

For the presentation of our results, we consider a selection of
four states---California (CA), North Carolina (NC), Indiana (IN)
and Mississippi (MS)---considered by Bell (\citeyear{bell99}). This
selection represents both small (i.e., large $V_i$) and large
(i.e., small $V_i$) states and thus should give us a fairly
general idea of the degree of accuracy of the Laplace and ADM
approximations with varying $V_i$ when compared to the exact
posterior distributions of the shrinkage factors obtained by
one-dimensional numerical integration.

First, consider the year 1993 when the REML esti\-mate of $A$ is
positive (1.7). The exact posterior distributions of the shrinkage
factors, ADM approximations and the  first-order Laplace
approximations (Kass and Steffey, \citeyear{ks89}) are plotted in
Figure~\ref{figsaipe93A}. The solid curves in Figures \ref{figsaipe93A} and \ref{figsaipe97A} are the
exact posterior distributions of $B_i$, which are obtained from
the posterior of $A$, under the prior, after multiplying by\break the
Jacobian and normalizing through numerical one-dimensional
integration. The dotted lines are first-order Laplace approximations
to the posterior distributions of $B_i$, which are
simply  normal distributions with means identical
to $B_i$ with $A$ replaced by its posterior mode and
variance expressions given in Kass and Steffy (\citeyear{ks89}).
Thus, the posterior means and variances of $B_i$ are essentially
approximated~by the first-order Laplace method.
From the plot it~is clear that the ADM approximation
is far better than the  first-order Laplace approximation when we
compare them with the exact posterior distribution of~$B_i$.

Table~\ref{tabcompariosnposteriormomentsB} displays the exact
posterior means and variances as well as their approximations for
these states. In general, the ADM approximation appears to be fairly accurate with
an indication of under-approximation of the posterior mean, especially for
states with small $V_i$ (CA, NC). On the other hand, the
first-order Laplace approximation  generally overestimates the
exact posterior means, sometimes by a large margin, and
approximates shrinkage factors for all the states in the year 1997
by 1. Turning to the posterior variances, we observe that the
first-order Laplace method generally over-approximates the exact
posterior variances, sometimes by a large margin, especially for
the year 1997. The poor performance of the Laplace method, for the
SAIPE 1997 data, can be attributed to the fact that the use of
uniform prior on $A$ yields a posterior mode that lies on the
boundary. The ADM approximation appears to perform well for both
the years, especially for 1997 when the Laplace method breaks
down. For the year 1993, the ADM method appears to slightly
under-approximate the exact posterior variances, especially for
the states with small $V_i$. Overall, it appears that the accuracy
of the ADM approximation depends somewhat on the states---the
larger the $V_i$ the better the approximation accuracy.

\begin{table}
\tabcolsep=0pt
  \caption{Comparison of posterior moments based on SAIPE state-level data}\label{tabcompariosnposteriormomentsB}
    \begin{tabular*}{\columnwidth}{@{\extracolsep{\fill}}lcccccc@{}}
    \hline
    & \multicolumn{3}{c}{\textbf{Posterior mean}} & \multicolumn{3}{c@{}}{\textbf{Posterior variance}} \\
\ccline{2-4,5-7}
    \textbf{State}    & \textbf{Exact} & \textbf{ADM}   & \textbf{Laplace}\tabnoteref{table11} & \textbf{Exact} & \textbf{ADM}   & \textbf{Laplace\tabnoteref{table11}}\\
\hline
\multicolumn{7}{@{}c@{}}{Results based on 1993 data} \\
    CA    & 0.47  & 0.37  & 0.56  & 0.038 & 0.023 & 0.093 \\
    NC    & 0.62  & 0.55  & 0.72  & 0.030 & 0.025 & 0.061 \\
    IN    & 0.80  & 0.77  & 0.87  & 0.014 & 0.014 & 0.019 \\
    MS    & 0.81  & 0.79  & 0.89  & 0.012 & 0.012 & 0.015 \\[5pt]
    \multicolumn{7}{@{}c@{}}{Results based on 1997 data} \\
    CA    & 0.68  & 0.60  & 1.00  & 0.037 & 0.041 & 0.987 \\
    NC    & 0.84  & 0.81  & 1.00  & 0.014 & 0.018 & 0.120 \\
    IN    & 0.87  & 0.85  & 1.00  & 0.010 & 0.013 & 0.071 \\
    MS    & 0.92  & 0.91  & 1.00  & 0.005 & 0.005 & 0.021 \\
\hline
\end{tabular*}
\tabnotetext[*]{table11}{Laplace first-order approximation; see Kass and Steffey (\citeyear{ks89}) for details.}
\end{table}%

We expect the accuracy of the Laplace approximation to depend on
the specific prior used for $A$.  In addition, the quality of both
first- and second-order Laplace approximations seems to depend on
$k$ and the $V_i/A$ values. For our SAIPE data analyses, we also
tried second-order Laplace approximations for both the years (not
reported here). The second-order Laplace approximation generally
improves on the accuracy for states with large $V_i$ when the
posterior mode is strictly positive.  However, when the posterior mode
is on the boundary (e.g., for the year 1997), the Laplace
second-order approximation produces undesirable results, such as
$\hat {B}_i>1$, negative posterior variance, etc. So we could not
even produce the graphs. For asymptotic expansions of posterior
expectations when the posterior mode is on the boundary, one might
need to consider approaches outlined in Erkanli (\citeyear{erkanli94}); this
can be tried in the future.  But even then we believe that  for
small~$k$ the Laplace method may not perform well in presence of
extreme skewness.

One important step in approximating the posterior distributions
used by Morris and Tang involves finding the ALM (Adjustment for
Likelihood Maximization) estimator of $A$ by maximizing the
product of the REML likelihood $L(A)$ and a universal adjustment
factor $A$ applicable to all the states primarily to avoid a zero
estimate of $A$. Given the above data analyses, is there any need
to find different adjustment factors, possibly depending on the
$V_i$, when approximating the posterior of $B_i$?

\section*{How May the ADM Method be Useful
in a Non-Bayesian Paradigm?}

While the method proposed in the paper under discussion is
essentially Bayesian with an innovative simple way to approximate
the exact Bayesian solution, one could use some of the ideas
presented in the paper in non-Bayesian approaches like the
empirical best linear unbiased prediction (EBLUP) widely used in
small area estimation. To elaborate on this point, first note that
the two-level model, given by~(1)--(2), can be viewed as the
following simple linear mixed model:
\[
y_i=x_i^{\prime}\beta + u_i + e_i,
\]
where $\{u_i\},$ area-specific random effects, and $\{e_i\},$
sampling errors, are independently distributed with $u_i\sim
N(0,A)$ and $e_i\sim N(0,V_i), i=1,\ldots,k$.
%

\begin{figure*}[b]

\includegraphics{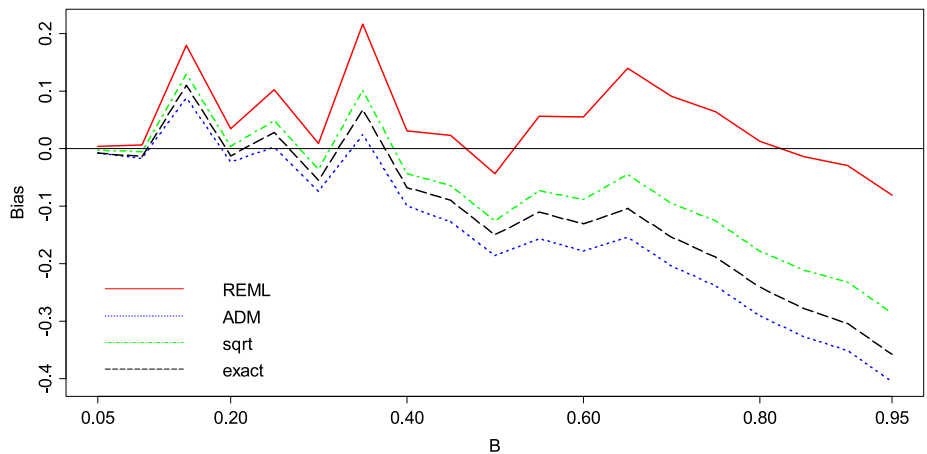}

\caption{Plot of simulated biases of different GML estimators of
$B_i$: the balanced case.} \label{figSimulBiasB}
\end{figure*}

The Bayes estimator of $\theta_i=x_i^{\prime}\beta + u_i$, as approximated by the ADM
method, is identical to an EBLUP of $\theta_i$ when the ALM
estimator of $A$ is used in place of REML, ML or other standard
variance component estimators.  The results on the frequentist
coverage (i.e., conditional on the hyperparameters $\beta$ and
$A$) of the approximate Bayesian intervals of~$\theta_i$ presented
in the Morris--Tang paper should be encouraging to the
non-Bayesians. However, from a~frequentist perspective, the
interesting problem of establishing the second-order accuracy of
coverage\break along the lines of Li and Lahiri (\citeyear{li2010adjusted}) remains open.
Morris and Tang suggested an approximation to the posterior
variance of $\theta_i$ as a measure of uncertainty of their point
estimate. However, since their point estimate of $\theta_i$  can
be viewed as an EBLUP, one may suggest the Morris--Tang measure of
uncertainty to estimate the traditional mean squared error (same
as the integrated Bayes risk, conditional on the hyperparameters)
as described in Jiang and Lahiri\break (\citeyear{jl06a}). It is not, however, clear
if the usual second-order unbiasedness criterion, advocated by the
non-Bayesians, would be satisfied by the approximate posterior
variance formula given in (58) of the Mor\-ris--Tang paper. We refer
the readers to Rao (\citeyear{rao03}) and Jiang and Lahiri (\citeyear{jl06a}) for a review of
the non-Bayesian methods.

The standard variance component estimation me\-thods such as the
REML and ML, despite their good asymptotic properties, frequently
yield zero estimates of the unknown variance component $A$.  This
is a lingering problem in the classical variance component
literature. For the model (1)--(2), the simulation results
given in Li and Lahiri (\citeyear{li2010adjusted}) suggest that the percentage of
zero estimates by the REML method depends on several factors,
including the variation of the ratios $V_i/A$ across the small
areas and the value of $k$. Li and Lahiri (\citeyear{li2010adjusted}) obtained an
adjusted maximum likelihood (AML) by multiplying the profile
likelihood, as given by $L_P(A)$ in Section 2 of Li and Lahiri (\citeyear{li2010adjusted}),
by an adjustment factor $A$. This translates to the following
adjustment factor:
\[
h(A)=A |X^{\prime}D^{-1}_{V+A}X|^{{1}/{2}}
\]
for  the corresponding residual likelihood, given in Section 2 of Li and Lahiri (\citeyear{li2010adjusted}).  Note that $h(A)$ differs from the
adjustment factor $A$ suggested in the paper under discussion by an additional\vspace*{2pt}
factor $|X^{\prime}D^{-1}_{V+A}X|^{{1}/{2}}$.

In the context of estimating the shrinkage factors~$B_i$, simulation results of
Li and Lahiri (\citeyear{li2010adjusted}) indicate lower biases of the shrinkage
estimators when  the Li--Lahiri adjustment factor is used, where
the bias is defined with respect to the marginal distribution of
$y$, given the hyperparameters $\b$ and~$A$.  In the context of a
general linear mixed model,  Lahiri and Li (\citeyear{lahirilifcsm2009}) proposed a
generalized maximum likelihood (GML) method, which includes ML,
REML, ALM and AML methods as special cases. For the model (1)--(2),
the GML essentially maximizes\break $h(A)\times L(A)$ with respect to
$A$, where $h(A)$ is a~general adjustment factor. This raises an
interesting question: how should one choose an adjustment factor
$h(A)$ in the GML method?

To fix ideas, we restrict ourselves to the class of adjustment
factors of the form $h(A)=A^q$. Since $\mathrm{V}(\hat B_i)$ is not
affected by $q$, up to the order $o(k^{-1})$
(Lahiri and Li, \citeyear{lahirilifcsm2009}), it makes sense to choose $q$ that
provides good properties in terms of the bias of the estimator. To
this end, using Lahiri and Li (\citeyear{lahirilifcsm2009}), we have
\[
\frac{\operatorname{Bias}(\hat B_i)}{\mathrm{V}(\hat B_i)}\approx \frac{1}{B_i}
\biggl(1-\frac{q}{1-B_i}  \biggr).
\]
Obviously,  $q=1-B_i$ is the ideal
choice---one that makes the bias/variance ratio nearly zero. While
we cannot use this choice since $A$ is unknown, it suggests the
range $[0,1]$ for  $q$.  Interestingly, the REML corresponds to
the choice $q=0$ while the Morris--Tang ALM corresponds to the other extreme
$q=1$. A~compromise choice is $q=0.5$, which corresponds to
$B_i=0.5$. In the Bayesian language, this choice would then
correspond to the prior $\pi (A)=1/\sqrt{A}$, a~prior  also
mentioned in the paper, since the Morris--Tang ADM recommends the
adjustment factor $A$ for any prior on $A$.  Figure 3 displays the
simulated biases of different estimators of the shrinkage factor
for the balanced version of model (1)--(2).  In terms of the bias,
the multiplier $\sqrt{A}$ usually works better than $A$.

Let us now explain how the ALM or AML method may help a
non-Bayesian method like the parametric bootstrap
(Chatterjee, Lahiri and Li, \citeyear{chatlahirili08}; Li and Lahiri, \citeyear{li2010adjusted}) in constructing
intervals  for the random effects $\theta_i$,
which requires repeated generation of a pivotal quantity from
several bootstrap samples. A strictly positive estimate of $A$ is
absolutely needed for this method since the pivotal quantity is
undefined when $A$ estimate is zero. A crude fix is to take a
small positive number whenever the estimate of $A$ is zero. But,
in a simulation study, Li and Lahiri (\citeyear{li2010adjusted}) observed that the
coverage errors and also the length of the parametric bootstrap
method could be sensitive to the choice of this positive
truncation point. The ALM or AML offers a sensible solution to
this problem of the parametric bootstrap method.
Li and Lahiri (\citeyear{li2010adjusted}) showed that the use of ALM or AML estimator
of $A$ improves on coverage as well as the length of the
parametric bootstrap interval estimate.

In the paper under discussion, Morris and Tang discuss the case of
a single variance parameter $A$.  Pramanik (\citeyear{pramanik2008bayesian})
extended the ADM method to the nested error regression model with
two unknown variance components by noting that one of the varian\-ce
components that corresponds to the within small area variation can
be integrated out.    However, it is not clear how the ADM method,
as proposed by Morris and Tang, would extend to a general linear
mixed model with more than two variance components, a situation
where a simple method such as the ADM method would be most
welcome.

We congratulate the authors for preparing an insightful and
informative paper on the ADM method.  This will surely inspire
others to contribute to this important area of research.

\section*{Acknowledgments}

We would like to thank Professor
Eric V. Slud, University of Maryland, College Park, for making a
number of constructive comments on an earlier version of our
discussion, and Dr. William R. Bell, U.S. Census Bureau, for some
useful discussion on the SAIPE data analyses. The first author's
research was supported in part by National Science Foundation
Grant SES-0851001, University of Michigan Contract 013448-001 and U.S. Census Bureau Contract
YA132309CN0057.

\vspace*{-2pt}

\end{document}